\documentclass[bib]{statapress}

\usepackage[crop,newcenter,frame]{pagedims}

\usepackage{amsmath}

\usepackage{amsmath, amssymb, amsfonts, booktabs}
\usepackage{bbm}

\usepackage{graphicx}

\usepackage{sj}

\usepackage{epstopdf}

\usepackage{epsfig}

\usepackage{stata}

\usepackage{shadow}

\sjsetissue{$\#$}{$\#$}{$\#$}{$2025$}

\usepackage{bm}

\begin{document}


\inserttype[]{article}

\author{Giovanni Cerulli}{
Giovanni Cerulli\\IRCrES-CNR\\Rome, Italy\\giovanni.cerulli@ircres.cnr.it
}

\title[Optimal Policy Learning for Multi-Action Treatment]
{Optimal Policy Learning for Multi-Action Treatment with Risk Preference using Stata}

\maketitle

\begin{abstract}
This paper presents the Stata community-distributed command \texttt{opl\_ma\_fb} (and the companion command \texttt{opl\_ma\_vf}), for implementing the \textit{first-best} Optimal Policy Learning (OPL) algorithm to estimate the best treatment assignment given the observation of an outcome, a multi-action (or multi-arm) treatment, and a set of observed covariates (features). It allows for different risk preferences in decision-making (i.e., risk-neutral, linear risk-averse, and quadratic risk-averse), and provides a graphical representation of the optimal policy, along with an estimate of the maximal welfare (i.e., the value-function estimated at optimal policy) using regression adjustment (RA), inverse-probability weighting (IPW), and doubly robust (DR) formulas.


\end{abstract}

\section[Introduction]{Introduction} \label{sec:introduction}

This paper presents the command \texttt{opl\_ma\_fb} (and the companion command \texttt{opl\_ma\_vf}), implementing the \textit{first-best} optimal policy learning (OPL) algorithm to estimate the best treatment assignment given the observation of an outcome, a multi-action (or multi-arm) treatment, and a set of observed covariates (or features). It allows for different risk preferences in decision-making (i.e., risk-neutral, risk-averse linear, risk-averse quadratic), and provide graphical representation of the optimal policy, along with an estimate of the maximal welfare (i.e., the value function estimated at the optimal policy).  

OPL is a fundamental methodology that, using counterfactual analysis and machine learning, determines the best decision strategy based on observational data. The objective is to learn a policy $\pi^*(X)$, with $X \in \mathcal{X}$, that maximizes the expected value of a given outcome of interest (i.e., the welfare), considering the available information on the units, in a setting where multiple actions $A \in \mathcal{A}$ are available. 

Optimal policy learning over a finite set of alternatives is a common problem across many domains, including finance, medicine, marketing, and public policy. In these settings, the objective is to identify the best possible action among several available ones, given a set of observed covariates (or features), in order to maximize a specific expected reward (or outcome). This task is formally addressed by seeking to estimate a decision rule - i.e. a policy - that maps feature configurations onto actions, so as to maximize the expected reward. The policy $\pi(X)$ assigns an action to each context $X$, and the optimal policy $\pi^*(X)$ is the one that maximizes the so-called value function (i.e., the average outcome or welfare).

This framework is general and finds practical applications in diverse domains. In medicine, personalized treatments seek to assign medical interventions (e.g., drugs, surgeries, therapies) to patients based on their clinical and personal characteristics, with the goal of maximizing recovery time or survival rates. In digital advertising, optimal allocation of customized ads relies on user behavior, preferences, and interaction history to maximize sales or user engagement. In finance, investment decisions such as stock selection can be cast as multi-action choices where the objective is to maximize capital gains while accounting for risk factors and market trends. In public policy, decision-makers may assign different types of financial support (e.g., grants, loans, tax credits) to firms or individuals in a way that maximizes long-term success or minimizes future vulnerability.

Across these domains, the OPL framework operates on a triplet data structure: 
(i) a signal from the environment (the features $X$); (ii) a multi-action treatment $A \in \{0,1,\dots,M{-}1\}$ where $M$ is the number of available actions/treatments, (iii) a reward $Y$ associated with the selected action/treatment.

The modeling approach presented in this paper can be framed within the literature of offline contextual multi-armed bandits (Auer et al., 2002; Slivkins, 2019; Silva et al., 2022), a class of reinforcement learning models in which each arm (action) has an unknown reward distribution conditional on the context. The objective is to learn, from observed data, which arm yields the highest expected reward. In classic online learning settings, this requires balancing exploration (trying new actions to learn their reward) and exploitation (choosing the current best-known action), giving rise to the well-known exploration-exploitation trade-off (Sutton and Barto, 2018; Li, 2023; Mui and Dewan, 2021).

In offline OPL with observational data, which is the one this paper refers to, the situation is different: a sufficiently large dataset already exists, containing past actions, features, and corresponding outcomes. Under assumptions of unconfoundedness and overlap, the learning task becomes a purely exploitative problem (Kitagawa and Tetenov, 2018; Athey and Wager, 2021; Zhou, Athey, and Wager, 2023; Bertsimas and Kallus, 2020). In this setting, the reward probabilities can be directly inferred from the data, and the optimal policy can be learned by maximizing empirical performance (Tschernutter, 2022; Wen and Li, 2023; Xin et al., 2020).

However, a limitation of standard OPL is its reliance on risk-neutral decision-making. Most methods assume that agents are indifferent to the variability of outcomes and only care about maximizing average returns. Yet, in many real-world contexts, especially in finance, health, and public policy, agents are risk-averse, as they prefer actions that yield more stable outcomes, even at the cost of lower expected values (Sani et al., 2012; Chandak et al., 2021; Cassel et al., 2023). 

Following Cerulli (2024), this Stata implementation extends the standard OPL framework to explicitly incorporate risk preferences, proposing decision rules that balance expected reward and outcome uncertainty. We provide in Stata an adjusted implementation of the first-best optimal policy, through the command \texttt{opl\_ma\_fb} (and the companion \texttt{opl\_ma\_vf}), which allows the user to define utility functions that reflect risk-neutral, linear risk-averse, or quadratic risk-averse preferences. These preferences are operationalized through the estimation of both conditional means and conditional variances of the outcome given action and features. The resulting decision rule selects the action that maximizes a utility function reflecting the agent's attitude toward risk.

Overall, this paper contributes to the empirical and computational literature on OPL by extending it to account for risk sensitivity in multi-action treatment settings, providing a flexible, data-driven tool for optimal decision-making under uncertainty using Stata. The method is general, applicable to many domains, and now can be fully operationalized through the proposed Stata commands for both training and evaluating optimal policies.

The paper is organized as follows. This section (section~\ref{sec:introduction}), outlines the motivation, objectives, and a general overview of the proposed methodology for optimal policy learning under risk preferences. Section~\ref{sec:value_policy} introduces the theoretical framework, defining the value function and the structure of the optimal policy, with particular attention to the role played by counterfactual outcomes. In Section~\ref{sec:estimators}, we describe the estimation strategies for the value function and highlight the methodological challenges associated with counterfactual inference. Section~\ref{sec:risk_preferences} extends the framework to incorporate risk preferences, detailing the implementation of risk-neutral, linear risk-averse, and quadratic risk-averse utility specifications. Sections~\ref{sec:implementation1} and \ref{sec:implementation2} provide implementation details in Stata through the \texttt{opl\_ma\_fb} and \texttt{opl\_ma\_vf} commands, including practical considerations about their implementation. Section~\ref{sec:applications} presents an empirical application to illustrate the effectiveness of the proposed methods. Finally, Section~\ref{sec:conclusion} concludes the paper with a discussion of key findings, limitations, and directions for future research.

\section{Value function and optimal policy} \label{sec:value_policy} 

Consider a decision maker who would like to allocate a set of $M$ different treatments over a given population based on individual characteristics, with the aim of maximizing a given measure of welfare (income, employability, recovery from a disease, etc.). 

For this decision maker, a policy $\pi(X)$ is defined as a decision rule, mapping individual characteristics onto one of $M$ alternative treatments (or actions), that is:

\begin{equation}
    \pi: X \longrightarrow \pi(X) = A \in \mathcal{A}
\label{eq:policy1}
\end{equation}   

For this decision maker, the value function (or \textit{welfare} measure), $V(\pi)$, is defined as the (average) capacity of a given policy $\pi$ in determining the expected (generally, non-negative) outcome $Y$, that is:

\begin{equation}
    V(\pi) = \mathbb{E} \left[ Y(\pi(X)) \right]
\label{eq:Vpi_multi}
\end{equation}
The \textit{optimal policy} $\pi^*$ is the one that maximizes this function:

\begin{equation}
    \pi^* = \arg\max_{\pi} \mathbb{E} \left[ Y(\pi(X)) \right]
\label{eq:pistar_multi}
\end{equation}
among all the policies $\pi \in \Pi$, where $\Pi$ is a set of feasible policies, possibly made of different classes. 

It is easy to see that the value function is  based on counterfactuals, as $Y(\pi(X))$ represents a potential outcome under action $\pi(X)$. Hence, counterfactual analysis is required to estimate both the effect of a chosen action and its maximum achievable welfare. 

For a multi-action setting where $A \in \mathcal{A} = \{0, 1, \dots, M-1\}$, by using the law of iterated expectation (LIE), we have that the expected mean outcome under the optimal policy $\pi^*(X)$ can be expressed as:

\begin{equation}
    \mathbb{E} \left[ Y(\pi^*(X)) \right] = \mathbb{E}_X \left[ \mathbb{E} \left[ Y(\pi^*(X)) \mid X \right] \right]
    \label{eq:iterated_expectation}
\end{equation}

which follows from the fact that: 

\begin{equation}
    \max_{\pi} \mathbb{E} \left[ Y(\pi(X)) \right] = \mathbb{E}_X \left[ \max_{a \in \mathcal{A}} \mathbb{E} \left[ Y(a) \mid X \right] \right]
\label{eq:maxpi_multi}
\end{equation}

The conditional expectation of the outcome given $X$ is then:

\begin{equation}
    \mathbb{E} \left[ Y(\pi(X)) \mid X \right] = \mathbb{E} \left[ Y(A) \mid X \right], \quad \text{for } A \in \mathcal{A}
\label{eq:expectation_multi}
\end{equation}

The \textit{first-best policy}, denoted \( \pi^{\text{FB}} \), is the optimal policy that assigns to each unit the action that maximizes the expected outcome, assuming full knowledge of the potential outcome functions. Formally:
\begin{equation}
\pi^{\text{FB}}(X) = \arg\max_{a \in \mathcal{A}} \mathbb{E}[Y(a) \mid X]
\end{equation}

This policy selects, for each value of \( X \), the action with the highest expected reward:
\[
\pi^{\text{FB}}(X) = \left\{ a : \max_{a' \in \mathcal{A}} \mu(a', X) \right\}, \quad \text{where } \mu(a, X) = \mathbb{E}[Y(a) \mid X]
\]

where the conditional average treatment effect (CATE) across multiple actions is:

\begin{equation}
    \tau(X, a, a') = \mathbb{E} \left[ Y(a) \mid X \right] - \mathbb{E} \left[ Y(a') \mid X \right], \quad \forall a, a' \in \mathcal{A}, a \neq a'
\end{equation}

The first-best solution is generally not identifiable in practice, as it requires counterfactual knowledge - i.e., the full knowledge of all potential outcomes \( Y(a) \) for all actions \( a \). The first-best corresponds to the optimal solution, which is only attainable in  settings without policy constraints (Bhattacharya and Dupas, 2012). In many policy contexts, such constraints typically restrict the policy space to specific classes, such as threshold-based, linear-combination, or decision-tree rules (Cerulli, 2023; 2025), which are however not considered in this paper.

\section{Estimators of the Value Function}
\label{sec:estimators} 

We saw that estimating the value function requires the knowledge of counterfactuals. For going ahead with estimation, we thus need to rely on two fundamental identification assumptions: (1) \textit{unconfoundedness}, and (2) \textit{overlapping}.   

The assumption of \textit{unconfoundedness} states that, conditional on observed covariates $X$, the treatment assignment $A$ is independent of the potential outcomes $Y(a)$:

\begin{equation}
    Y(a) \perp A \mid X, \quad \forall a \in \mathcal{A}
\end{equation}

This assumption is crucial because it ensures that any observed differences in outcomes across different actions $a \in \mathcal{A}$ are attributable to the treatment itself rather than unobserved confounders. Importantly, this assumption makes it possible to express counterfactuals in terms of observable components, specifically: 

\begin{equation}
    \mathbb{E}[Y(a) \mid X]=\mathbb{E}[Y \mid X, A = a]
\label{eq:equationUnconf}
\end{equation}
which makes counterfatuals estimable via an observable conditional expectation. This is a fundamental equation in OPL.   

The assumption of \textit{overlapping} (or \textit{positivity}) requires that each action has a strictly positive probability of being assigned to any unit, conditional on $X$:

\begin{equation}
    0 < P(A = a \mid X) < 1, \quad \forall a \in \mathcal{A}, \forall X
\end{equation}

This assumption ensures that for every possible value of $X$, we observe units that receive each action in $\mathcal{A}$. If this assumption is violated (e.g., if only certain groups never receive a particular treatment), we lack empirical support to estimate the counterfactual outcome under that action, leading to unreliable policy learning.

Under \textit{unconfoundedness} and \textit{overlapping}, three standard estimators of the value function have been proposed in the literature: the Regression Adjustment (RA), the Inverse Probability Weighting (IPW), and the Doubly Robust (DR) estimators. Each of these methods has advantages and limitations (Dudik et al., 2011), and their appropriateness depends on the nature of the data and the assumptions made about the treatment assignment process.

\subsection{Regression adjustment (RA)}
The RA (sometimes referred to as the \textit{direct method}) estimates the value function using regression estimates of the counterfactual (potential) outcomes. The RA formula is:  

\begin{equation} \label{eq:regAdj}
\hat{V}_{RA}(\pi) = \frac{1}{N}  \sum_{i=1}^{N}  \hat{Q}_{i}(X_{i},\pi(X_{i})) 
\end{equation}
where $\hat{Q}_{i}(X_{i},\pi(X_{i}))=
\sum_{a=0}^{M-1} \hat{Q}_{i}(X_{i},a) \cdot \pi_{ia}$, with 
$\pi_{ia}=1[\pi_{i}=a]$. The RA approach provides a consistent estimation of the value function provided that the functional form of the regression model $\hat{Q}(X, A=a)$ is a consistent estimation of $\mathbb{E}(Y|X,A=a)$.
This estimator is straightforward to implement and can be efficient when the model is correctly specified.
However, it is highly sensitive to misspecification: if the regression model does not accurately capture the true relationship between $X$, $A$, and $Y$, the estimation of the value function can be unreliable.

\subsection{Inverse Probability Weighting (IPW)}
The IPW estimator corrects for selection bias by re-weighting observations according to the probability of receiving a given treatment:
\begin{equation}
    \hat{V}_{IPW}(\pi) = \frac{1}{N} \sum_{i=1}^{N}  \left[ \frac{\mathbbm{1}(A_i = \pi(X_i)) Y_i}{\hat{P}(A_i \mid X_i)} \right]
\end{equation}
where $\hat{P}(A_i \mid X_i)$ is the estimated propensity score, representing the probability of receiving treatment $A$ given $X$.

IPW ensures an unbiased estimate of the value function under correct specification of the propensity score model. By weighting observations according to their inverse probability of treatment assignment, IPW effectively balances treated and untreated groups as if treatment were randomly assigned. However, this approach has some drawbacks: (i) it can be inefficient if the estimated propensity scores are too close to 0 or 1, leading to extreme weights and high variance; (ii) it requires a well-specified treatment assignment model; if $P(A \mid X)$ is misspecified, IPW will produce biased estimates; (iii) it is particularly sensitive to violations of the overlapping assumption, as observations with low probability of treatment assignment receive excessively high weights, causing instability in estimation.

\subsection{Doubly Robust (DR) Estimator}
The DR estimator combines both the RA and IPW to achieve robustness against model misspecification:
\begin{equation}
    \hat{V}_{DR}(\pi) = \frac{1}{N}  \sum_{i=1}^{N}  \left[ \hat{Q_i}(X_i, \pi(X_i)) + \frac{\mathbbm{1}(A = \pi(X_i))(Y - \hat{Q}_i(X_i, A_i))}{\hat{P}(A_i \mid X_i)} \right]
\end{equation}
This estimator remains consistent if either the outcome model $Q(X, A)$ or the propensity score model $P(A \mid X)$ is incorrectly specified.

The key advantage of DR estimation is its robustness: it remains unbiased as long as at least one of the two models (outcome regression or propensity score) is correctly specified. This property makes it particularly useful when there is uncertainty about model specification. However, DR estimation also has some limitations: (i) if both the outcome model and the propensity score model are misspecified, DR estimates will still be biased; (ii) the method requires careful implementation, as estimation errors in both models can compound and affect stability; (iii) in cases where propensity scores are close to 0 or 1, the weighting term in DR can still suffer from high variance, similar to IPW. Despite these challenges, the doubly robust estimator is often preferred because it offers greater protection against misspecification errors than either RA or IPW alone. It is particularly useful in observational studies where treatment assignment mechanisms and outcome relationships are uncertain.

\section{Optimal decision under reward uncertainty}\label{sec:risk_preferences} 
In an uncertain environment, the returns from undertaking specific actions are associated to risk and uncertainty. In such a context, choosing, let's say, action A instead of action B depends not only on the average return of each option, but also on the uncertainty in obtaining such return. Therefore, decision-making must ponder the return and its related variability.   

\begin{figure}[t]
\centering
\includegraphics[width=11cm]{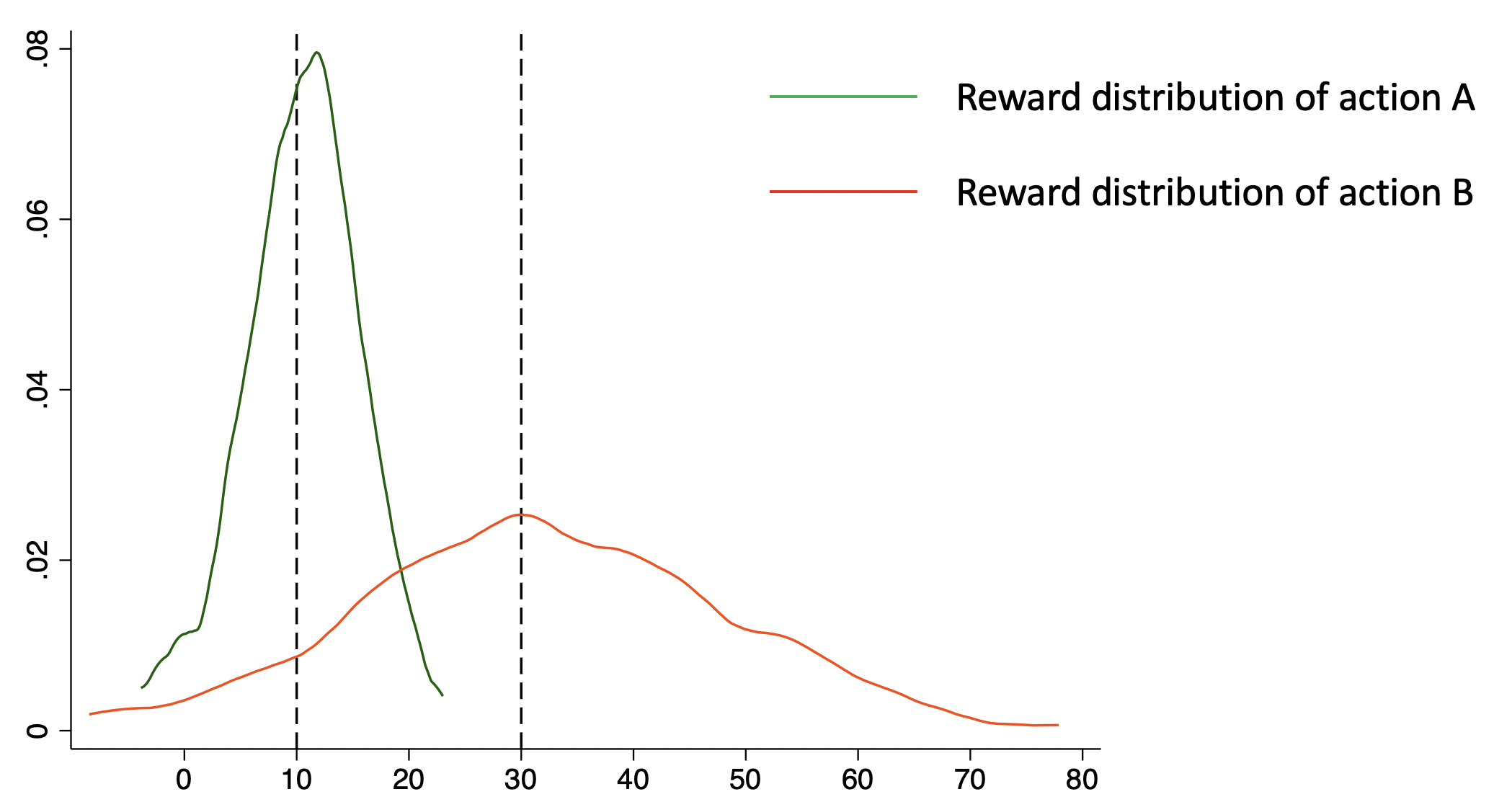}
\caption{Reward distribution and uncertainty realtive to two action, A and B. Action A provides a lower average return, but with smaller uncertainty. Action B provides a higher average return, but with larger uncertainty.}
\label{fig:fig1}
\end{figure}

Figure \ref{fig:fig1} shows the reward distribution and related uncertainty for two actions, A and B. We see that action A provides a lower average return, but with smaller uncertainty, whereas action B provides a higher average return but with larger uncertainty. In this case, it is not clear what action should be optimally undertaken, as a trade-off between expected reward and uncertainty takes place.

The issue has been well-recognized by a recent stream of multi-armed bandit literature focusing on risk-averse agents taking decisions not only on the basis of average reward, but also incorporating reward's uncertainty in their choice measured using, for example, the variance of the reward distribution (Sani et al., 2012). When the objective function incorporates risk, traditional algorithms trading-off exploration and exploitation with the aim of minimizing the policy regret, can take a different form and can have different asymptotic performance compared to traditional risk-neutral algorithms. 

In OPL with observational data (Sani et al., 2012; Chandak et al.; 2021; Cassel et al.; 2023), scholars aim to estimate the overall variance of the policy. For example, Chandak et al. (2021) provide consistent estimation of the offline variance of the return associated to the policy $\pi$ defined as:
\begin{equation} \label{eq:var1}
\sigma^{2}(\pi) = \text{Var}[Y(\pi(X)]
\end{equation}
Indeed, the return distribution is not only characterized by a central measure like the average reward of equation (\ref{eq:Vpi_multi}), but also by variability around this central measure.

Cerulli (2024) proposes a pretty different approach. Instead of focusing on the estimation of the overall total variance of the outcome related to a certain policy (that is, Eq. (\ref{eq:var1})), he proposes to focus on the estimation of the conditional variance, and introduce specific risk preferences. Let's delve into this approach. 

Conditional uncertainty can be measured via the conditional variance, which is the variance of the distribution of $Y|X$. The formula of the conditional variance is:
\begin{equation} \label{eq:variance}
\text{Var}(Y|X) = E[Y – \text{E}(Y)|X]^2 = \text{E}(Y^2|X) – {\text{E}(Y|X)}^2 
\end{equation}
We proceed action-wise. Therefore, for observation $i$, we can estimate the conditional variance associated to action $a$ as:
$$\sigma^{2}_{i}(a,X_{i}) = \text{Var}(Y_{i}|A_{i}=a,X_{i})$$
which can be easily estimated as the difference between two conditional means, similarly to formula (\ref{eq:variance}):
\begin{equation} \label{eq:variance2}
\hat{\sigma}^{2}_{i}(a,X_{i}) = \hat{\text{E}}(Y^{2}_{i}|A_{i}=a,X_{i}) - \hat{\text{E}}(Y_{i}|A_{i}=a,X_{i})^2
\end{equation}
where the conditional means in the RHS can be estimated using specific machine learning techniques. Thus, the optimal action to select, given the signal $X_{i}$, depends on the return/risk pair:
$$ [ \hat{\mu}_{i}(a,X_{i}) , \hat{\sigma}_{i}(a,X_{i})] $$
and preferences over them. Observe that $\hat{\sigma}_{i}(\cdot)$ is the estimated standard deviation. 

If we assume a risk-neutral setting, we state that people are indifferent to risk, and decisions are taken just looking at the maximum return. This is the standard model, where no uncertainty is considered.

On the contrary, if we assume a \textit{risk-averse} setting, i.e. one where people prefer lower levels of risk for a given level of return, we are introducing risk-preferences. A utility function for a risk-averse decision-maker would reflect this preference by assigning a lower utility value to actions with higher levels of risk.
Risk-averse preferences can be modeled through a utility function whose arguments are the conditional average reward and the conditional standard deviation. Here, we consider two settings: (i) \textit{linear} risk-averse preferences; and (ii) \textit{quadratic} risk-averse preferences. Importantly, two distinct actions can have a different preferential ordering according to the specific type of preferences assumed (Cerulli, 2024).
\\
\\
\textit{Linear risk-averse preferences}. The utility function is equal to the ratio between the conditional average reward and the conditional standard deviation:  
\begin{equation} \label{eq:linear_pref1}
U_{i,L}=\frac{\hat{\mu}_{i}}{\hat{\sigma}_{i}}
\end{equation}
implying, by equalizing $U_{i,L}$ to a constant $k$, a linear indifference curve:
\begin{equation} \label{eq:linear_pref2}
\hat{\mu}_{i} = k \cdot \hat{\sigma}_{i}
\end{equation}
\\
\\
\textit{Quadratic risk-averse preferences}. The utility function is equal to the ratio between the conditional average reward and the squared value of the conditional standard deviation:
\begin{equation} \label{eq:linear_pref}
U_{i,Q}=\frac{\hat{\mu}_{i}}{\hat{\sigma}_{i}^{2}}
\end{equation}
implying, by equalizing $U_{i,Q}$ to a constant $k$, a quadratic indifference curve:
\begin{equation} \label{eq:linear_pref}
\hat{\mu}_{i} = k \cdot \hat{\sigma}_{i}^{2}
\end{equation}
We can build examples of actions' preferential ordering where, according to linear risk-averse preferences, an agent turns out to prefer action A over action B, while according to quadratic risk-averse preferences, an agent is indifferent between action A and B, or even prefer B over A. 

We can conclude that, when comparing alternative actions under different risk-averse preferences, the action preferential ordering can change, and treatment allocation significantly differ. This leads to select a policy rule that, by weighting average returns according to their risk, does not coincide with the first-best risk-neutral policy, thus being sub-optimal compared to this benchmark. This produces a regret, i.e. a loss of welfare, that one can interpret as a loss due to a prudent attitude toward risk.     

For a linear risk-aversion (LRA) setting (and, similarly, for a quadratic one), the selected policy rule becomes:

\begin{equation} \label{eq:LRApolicy1}
\pi^{\text{LRA}}(X) = \left\{ a : \max_{a' \in \mathcal{A}} \frac{\mu(a', X)}{\sigma(a', X)} \right\}
\end{equation}
with:
\begin{equation} \label{eq:LRApolicy2}
V(\pi^{\text{LRA}}(X)) \leq V(\pi^{\text{FB}}(X))
\end{equation}
and regret:
\begin{equation} \label{eq:LRApolicy2}
R = V(\pi^{\text{FB}}(X)) - V(\pi^{\text{LRA}}(X)).
\end{equation}
It is thus intriguing to explore the extent to which different risk settings can influence the optimal actions selected and the corresponding regret. 

Assuming conditional independence, our Stata implementation of the first-best policy learning in a multi-action setting incorporates both linear and quadratic risk-adjusted preferences using the first-best rule as reference (optimal) decision algorithm.

\section{Syntax}

\subsection{Syntax for \texttt{opl\_ma\_fb}} \label{sec:syntax1}
\label{sec:implementation1} 

The command \texttt{opl\_ma\_fb} implements  Optimal Policy Learning (OPL) in a multi-action treatment setting computing the \textit{first-best} policy using the RA approach for estimating the value function under different risk preferences. It allows for training and evaluation of optimal treatment policies based on observational data. This command uses linear regression for estimating nuisance conditional means. The syntax is:

\begin{stsyntax}
\dunderbar{}opl\_ma\_fb
    \depvar\
    \varlist\  ,    
    {\tt policy\_train({\it varname\/})}
    {\tt model({\it string\/})}
    {\tt name\_opt\_policy({\it name\/})}
    \optional{
    {\tt match\_name({\it name\/})}
    {\tt new\_data({\it name\/})}
    {\tt policy\_non\_optimal\_train({\it varname\/})}
    {\tt policy\_non\_optimal\_new({\it varname\/})}
    {\tt value\_var({\it number\/})}
    {\tt save\_preds\_vars({\it name\/})}
    {\tt gr\_action\_train({\it name\/})}
    {\tt gr\_reward\_train({\it name\/})}
    {\tt gr\_reward\_new({\it name\/})}
    }
\end{stsyntax}

\noindent 
where:

\hangpara
\textit{depvar} is a numerical variable representing the outcome (or reward) of interest. It is a non-negative variable.

\hangpara
\textit{varlist} is a list of numerical variables representing the predictors, including categorical variables.

\vspace{0.5cm}

\textbf{Main options}

\hangpara
{\tt policy\_train({\it varname\/})} specifies the training variable for policy estimation.

\hangpara
{\tt model({\it string\/})} specifies the decision model, that can be one of: \texttt{risk\_neutral}, which considers only expected reward (no variance or risk are accounted for); \texttt{risk\_averse\_linear}, which adjusts reward by a linear function of its variance; and \texttt{risk\_averse\_quadratic}, which adjusts reward by a quadratic function of its variance.

\hangpara
{\tt name\_opt\_policy({\it name\/})} assigns a name to the optimal policy.

\vspace{0.5cm}

\textbf{Optional options}

\hangpara
{\tt match\_name({\it name\/})} specifies the name of the binary variable (0/1) that stores whether the actual treatment matches the optimal one.

\hangpara
{\tt new\_data({\it name\/})} provides a second dataset to predict optimal actions for new units. This dataset must contain the same features' names as the training dataset. 

\hangpara
{\tt policy\_non\_optimal\_train({\it varname\/})} is an alternative (non-optimal) policy to compare against training policy within training data.

\hangpara
{\tt policy\_non\_optimal\_new({\it varname\/})} is an alternative (non-optimal) policy to compare against optimal policy within new data.

\hangpara
{\tt value\_var({\it number\/})} imputes to a value equal to {\it number\/} possible negative values of the estimated conditional variance. 

\hangpara    
{\tt save\_preds\_vars({\it name\/})} saves conditional expectations and variances, at individual levels, in a dataset named {\it name\/}, saved in the user's current directory. 

\hangpara
{\tt gr\_action\_train({\it name\/})} generates a graph comparing actual vs. optimal action allocation in the training dataset. 

\hangpara
{\tt gr\_reward\_train({\it name\/})} generates a graph comparing actual vs. maximal expected reward in the training dataset.

\hangpara
{\tt gr\_reward\_new({\it name\/})} generates a graph showing maximal expected reward for new policy observations.

\vspace{0.5cm}

\textbf{Returns}

\hangpara
{\tt e(N\_train)} is the number of observations in the training dataset.

\hangpara
{\tt e(N\_new)} is the number of observations in the new (unlabeled) dataset.

\hangpara
{\tt e(N\_train\_opt\_pol)} is the number of observations for computing the optimal policy in the training dataset.

\hangpara
{\tt e(V\_train)} is the value function in the training dataset.

\hangpara
{\tt e(N\_V\_train)} is the number of observations for computing the value function in the training dataset.

\hangpara
{\tt e(V\_non\_opt\_train)} is the value function in the training dataset for the non-optimal policy.

\hangpara
{\tt e(N\_V\_non\_opt\_train)} is the number of observations for computing the value function in the non-optimal training dataset.

\hangpara
{\tt e(V\_opt\_train)} is the value function with optimal policy in the training dataset.

\hangpara
{\tt e(N\_V\_opt\_train)} is the number of observations for computing the value function in the training dataset with optimal policy.

\hangpara
{\tt e(V\_opt\_new)} is the value function in the new dataset for the optimal policy.

\hangpara
{\tt e(N\_V\_opt\_new)} is the number of observations for computing the value function in the new dataset for the optimal policy.

\hangpara
{\tt e(rate\_opt\_match)} is the rate of matches between the optimal and the training policy.

\vspace{0.5cm}

\textbf{Generated variables}

\hangpara 

\texttt{\_index}: indicator variable specifying the dataset source of each observation (0 = training data; 1 = new data).

\texttt{\_opt\_policy}: the estimated optimal policy rule, assigning to each unit the treatment that maximizes expected welfare.  

\hangpara 
\texttt{\_Y\_hat\_policy\_train}: the predicted outcome under the actual (observed) training policy, i.e. the historical assignment rule applied in the data.  

\hangpara 
\texttt{\_Y\_hat\_policy\_train\_non\_optimal}: the predicted outcome under a given non-optimal policy provided in the training set, used as a benchmark for comparison.  

\hangpara 
\texttt{\_Y\_hat\_policy\_optimal}: the predicted outcome under the estimated optimal policy, i.e. the counterfactual outcome distribution if all units had followed the first-best policy.  

\hangpara 
\texttt{\_match\_var}: an indicator variable equal to 1 if the actual treatment coincides with the estimated optimal treatment, and 0 otherwise. It measures the rate of alignment between historical and optimal assignments.  

\vspace{0.5cm}

\textbf{Installation}

\hangpara
For installing this command, consider to type:

\begin{stverbatim}
\begin{verbatim}
. ssc install opl_ma_fb
\end{verbatim}
\end{stverbatim}

\subsection{Syntax for \texttt{opl\_ma\_vf}} \label{sec:syntax2}
\label{sec:implementation2} 

The command \texttt{opl\_ma\_vf} estimates the value function for multi-action optimal policy learning using three different methods:

\begin{itemize}
    \item \textit{Regression Adjustment} (RA): estimates potential outcomes for each action using regression models.
    \item \textit{Inverse Probability Weighting} (IPW): uses estimated propensity scores to reweigh observations.
    \item \textit{Doubly Robust} (DR): combines RA and IPW for a more robust estimator.
\end{itemize}

This command uses linear regression for estimating nuisance conditional means. The syntax is:

\begin{stsyntax}
\dunderbar{}opl\_ma\_vf
    \depvar\
    \varlist\  ,    
    {\tt policy\_train({\it varname\/})}
    {\tt policy\_new({\it varname\/})}
\end{stsyntax}

\noindent 
where:

\hangpara
\textit{depvar} is a numerical variable representing the outcome of interest.

\hangpara
\textit{varlist} is a list of numerical variables representing the predictors, including categorical variables.

\vspace{0.5cm}

\textbf{Main options}

\hangpara
{\tt policy\_train({\it varname\/})} specifies the treatment policy variable used in training.

\hangpara
{\tt policy\_new({\it varname\/})} specifies the new policy variable to be evaluated.

\vspace{0.5cm}

\textbf{Returns}

\hangpara 
\texttt{e(N\_obs)}: Number of observations in the training dataset.

\hangpara  
\texttt{e(RA)}: Estimated value function using Regression Adjustment.

\hangpara 
\texttt{e(IPW)}: Estimated value function using Inverse Probability Weighting.

\hangpara 
\texttt{e(DR)}: Estimated value function using the Doubly Robust method.

\vspace{0.5cm}

\textbf{Installation}

\hangpara
For installing this command, consider to type:

\begin{stverbatim}
\begin{verbatim}
. ssc install opl_ma_vf
\end{verbatim}
\end{stverbatim}

\subsection{Syntax for \texttt{opl\_best\_treat}} \label{sec:syntax3}

The command \texttt{opl\_best\_treat} is a utility command to be used after running \texttt{opl\_ma\_fb} and loading the dataset stored into the option {\tt save\_preds\_vars({\it name\/})} which saves conditional expectations and variances, at individual levels, in a new dataset named {\it name\/}. The syntax is:

\begin{stsyntax}
\dunderbar{}opl\_best\_treat
    \varlist\
\end{stsyntax}
where \texttt{varlist} are the reward counterfactual predictions obtained after estimating \texttt{opl\_ma\_fb}, that is, \texttt{\_\_pred0}, \texttt{\_\_pred1}, \texttt{\_\_pred3}, $\dots$.  

The returns of this command are two variables: \texttt{\_\_Y\_hat\_max}, the maximal outcome; and \texttt{\_\_T\_best} the best treatment, according to the first-best decision rule.

\subsection{Syntax for \texttt{opl\_plot\_best}} \label{sec:syntax4}

The command \texttt{opl\_plot\_best} is a utility command to be used after running \texttt{opl\_best\_treat} to plot the observed versus the maximal expected reward, as well as the observed and optimal treatments. The syntax is:

\begin{stsyntax}
\dunderbar{}opl\_plot\_best
    \varlist\ , 
    \optional{
    {\tt gr\_reward\_train({\it name\/})}
    {\tt gr\_action\_train({\it name\/})}
    }
\end{stsyntax}
where \texttt{varlist} must be an ordered list of these four variables:
(first) the expected outcome (based on the training policy);
(second) the observed treatment (i.e., the training policy);
(third) the maximal expected outcome (based on the best policy);
(forth) the best treatment (i.e., the optimal policy).

The two optional options are used for saving the two plots in the user's current directory.  

\section{Application} \label{sec:applications} 

In this application, we aim to demonstrate how to estimate and evaluate an optimal policy using observational data from a simulated educational setting. The analysis relies on the \texttt{opl\_ma\_fb} command. The goal is to train a data-driven policy that maps student characteristics to one of three educational interventions differing by test difficulty levels with the aim of maximizing student performance.

We use the \texttt{spmdata} dataset, which contains simulated data from a hypothetical scenario designed to illustrate how to learn an optimal policy rule $\pi^{*}(X)$ under a given training policy setting. This dataset was originally introduced by Cattaneo, Drukker, and Holland (2013).

The training policy is as follows: at the start of the school year, a policy decision $A \in \{0, 1, 2\}$ is made for each student, representing the assigned level of test difficulty in their classroom:

\begin{itemize}
\item \texttt{A} = 0: classes with standard tests,
\item \texttt{A} = 1: classes with tests made of hard questions,
\item \texttt{A} = 2: classes with tests made of even harder questions.
\end{itemize}

At the end of the year, all students take a common evaluation test, independent of the initial treatment, resulting in a performance outcome index, \texttt{spmeasure}, constructed from both test scores and interview data. The learning goal is to determine whether the actual assignment to treatment $A$ was or wasn't optimal in the sense of maximizing the overall welfare, measured by the estimated value-function.  

In this setting, we assume that potential outcomes $Y(A)$ are conditionally independent of treatment assignment $A$ given observed covariates, which include \texttt{pindex} (parental status) and \texttt{eindex} (student’s environmental index). This conditional independence assumption justifies the application of observational OPL methods under the assumption of selection-on-observables.

We start by computing the first-best policy and value function using RA in a risk-neutral setting. Below the Stata code. 

\begin{verbatim}

********************************************************************************
* COMPUTING FIRST-BEST POLICY AND VALUE FUNCTION IN A RISK-NEUTRAL SETTING
********************************************************************************
* Load the dataset
sysuse spmdata , clear

* Rescale outcome variable for interpretability
replace spmeasure = spmeasure * 100

* Remove invalid observations
drop if spmeasure <= 0

* Ensure clean ID space and generate small test subsample
cap drop id
set seed 1010
sample 3
gen id = _n

* Define global macros for outcome, treatment, and covariates
global y "spmeasure"
global A "w"
global X "pindex eindex"

* Split the data into training and evaluation (new) sets
splitsample, gen(split, replace) nsplit(2) rseed(1010)

********************************************************************************
* STEP 1: Create synthetic non-optimal policy for NEW data (split == 2)
********************************************************************************
preserve
keep if split == 2                  // keep only the evaluation set
keep $X                             // keep only the covariates

* Generate uniform random numbers for random policy assignment
gen u = uniform()
gen treat_non_optimal_new = .

* Assign non-optimal policy randomly (uniformly across actions)
replace treat_non_optimal_new = 0 if u >= 0 & u < 0.33
replace treat_non_optimal_new = 1 if u >= 0.33 & u < 0.66
replace treat_non_optimal_new = 2 if u >= 0.66 & u < 1

* Save the new dataset
save my_new_data.dta, replace
restore

********************************************************************************
* STEP 2: Prepare the training set (split == 1) and simulate a non-optimal policy
********************************************************************************
keep if split == 1

* Generate non-optimal (random) policy for training set
gen u = uniform()
gen treat_non_optimal_train = .
replace treat_non_optimal_train = 0 if u >= 0 & u < 0.33
replace treat_non_optimal_train = 1 if u >= 0.33 & u < 0.66
replace treat_non_optimal_train = 2 if u >= 0.66 & u < 1

* Save the dataset as "mydata"
save mydata , replace
********************************************************************************
\end{verbatim}

We begin by loading the dataset \texttt{spmdata}. The outcome variable, \texttt{spmeasure}, captures student performance and is rescaled to enhance interpretability. We eliminate observations with zero or negative performance scores, as they may indicate data errors or violate the modeling assumptions, especially when calculating the value-function.

To ensure reproducibility and allow for individual tracking, we reset any existing individual identifier variable and assign a new one. For demonstration purposes, we extract a small subsample to facilitate initial testing of the workflow. Global macros are defined to simplify the use of variables throughout the script: the outcome of interest (\texttt{spmeasure}), the treatment indicator (\texttt{w}), and the covariates (\texttt{pindex} and \texttt{eindex}), which represent parental and environmental characteristics.

The core of the OPL approach involves learning and evaluating treatment policies. To mimic real-world decision-making settings, because we do not have a real dataset for new individuals, we randomly divide the initial dataset into two parts: a ``training'' set, used to learn the optimal policy, and a ``new'' set, used to project the optimal policy to new units (this split is controlled through a random seed to ensure replicability).

In Step 1, we prepare the evaluation set (\texttt{split=2}). Here, we retain only the covariates, simulating a realistic scenario where we must recommend a treatment without yet observing the outcome. We then generate a synthetic, non-optimal policy by randomly assigning actions with equal probability across the three alternatives. This baseline policy will serve as a benchmark when comparing the learned optimal policy’s performance.

Step 2 mirrors this logic within the training set. We generate a comparable non-optimal policy, again using random assignment, that can be used for in-sample comparisons against the estimated optimal policy.

Step 3 is the core of this application. Its code and results are listed below.

\begin{verbatim}
********************************************************************************
* STEP 3: Estimate the Optimal Policy using opl_ma_fb
********************************************************************************
global model "risk_neutral"

opl_ma_fb $y $X, ///
    policy_train($A) ///
    model($model) ///
    name_opt_policy("_opt_policy") ///
    match_name("_match_var") ///
    new_data(my_new_data) ///
    gr_action_train("gr1") ///
    gr_reward_train("gr2") ///
    gr_reward_new("gr3") ///
    save_preds_vars(my_results) ///
    policy_non_optimal_train(treat_non_optimal_train)
-------------------------------------------------------
MAIN RESULTS
-------------------------------------------------------
--> Data information
-------------------------------------------------------
Number of training observations = 56
Number of used training observations (optimal policy) = 56
Number of used training observations (non-optimal policy) = 56
Number of new observations = 56
Number of used new observations (optimal policy) = 56
Number of used new observations (non-optimal policy) = .
-------------------------------------------------------
--> Policy information
-------------------------------------------------------
Target variable: spmeasure
Features:  pindex eindex
Policy variable: w
Number of actions: 3
Actions: {0 1 2}
-------------------------------------------------------
Frequencies of the actions in the training dataset
-------------------------------------------------------

Multivalued |
 treatment: |
    j=0,1,2 |      Freq.     Percent        Cum.
------------+-----------------------------------
          0 |         34       60.71       60.71
          1 |         13       23.21       83.93
          2 |          9       16.07      100.00
------------+-----------------------------------
      Total |         56      100.00
-------------------------------------------------------
-------------------------------------------------------
--> Training data
-------------------------------------------------------
Value-function of the policy (training) = 68.95
Value-function of the non-optimal policy (training) = 57.97
Value-function of the optimal policy (training) = 124.19
Rate of optimal policy matches = .52
-------------------------------------------------------
--> New data
-------------------------------------------------------
Value-function of the non-optimal policy (new) = .
Value-function of the optimal policy (new) = 139.97
-------------------------------------------------------
\end{verbatim}

We set out by invoking the \texttt{opl\_ma\_fb} command to estimate the optimal policy under a \textit{risk-neutral} assumption. The model uses the training data to learn which actions lead to the highest expected outcomes, conditional on the covariates. We pass both the training and new data, the non-optimal benchmark policies, and optional structures for action and reward groupings. We also save predicted expectations and variances for further inspection or visualization.

At this point, the optimal policy has been learned and can be compared against random alternatives in terms of expected value. Additional evaluation (e.g., using \texttt{opl\_ma\_vf}) could further quantify how much welfare is gained by implementing the learned policy over sub-optimal ones.

The results from running \texttt{opl\_ma\_fb} provide a comprehensive picture of how the learned optimal policy performs relative to a baseline random (non-optimal) policy, both on the training data and on a new evaluation set.

The training sample consists of 56 observations, evenly used for both the optimal and non-optimal policy evaluations. Each student is assigned to one of three possible treatment levels (test difficulty), with the majority (60.71\%) receiving the standard version ($A$ = 0). This imbalance reflects real-world tendencies, where certain interventions (e.g., standard tests) are more common than others.

The learned optimal policy achieves a value function of 124.19 on the training set, compared to 57.97 for the randomly generated non-optimal policy, and 68.95 for the actual observed (historical) policy. This substantial improvement, more than doubling the value function compared to the benchmark, demonstrates the power of policy learning: tailoring decisions based on covariates (\texttt{pindex}, \texttt{eindex}) can significantly enhance outcomes.

Interestingly, the rate of optimal policy matches is only 0.52, suggesting that the learned optimal policy deviates considerably from the observed policy. This reinforces the idea that historical assignments were likely sub-optimal, and that a model-driven policy can provide superior guidance.

When applied to new data (unseen during training), the learned optimal policy maintains strong performance, reaching a value-function of 139.97. Observe that the value of the non-optimal policy on the new data is not available in this run of the command, but can be provided in a possible second run, upon inputting a specific non-optimal policy. 

\begin{figure}[t]
\centering
\includegraphics[width=15cm]{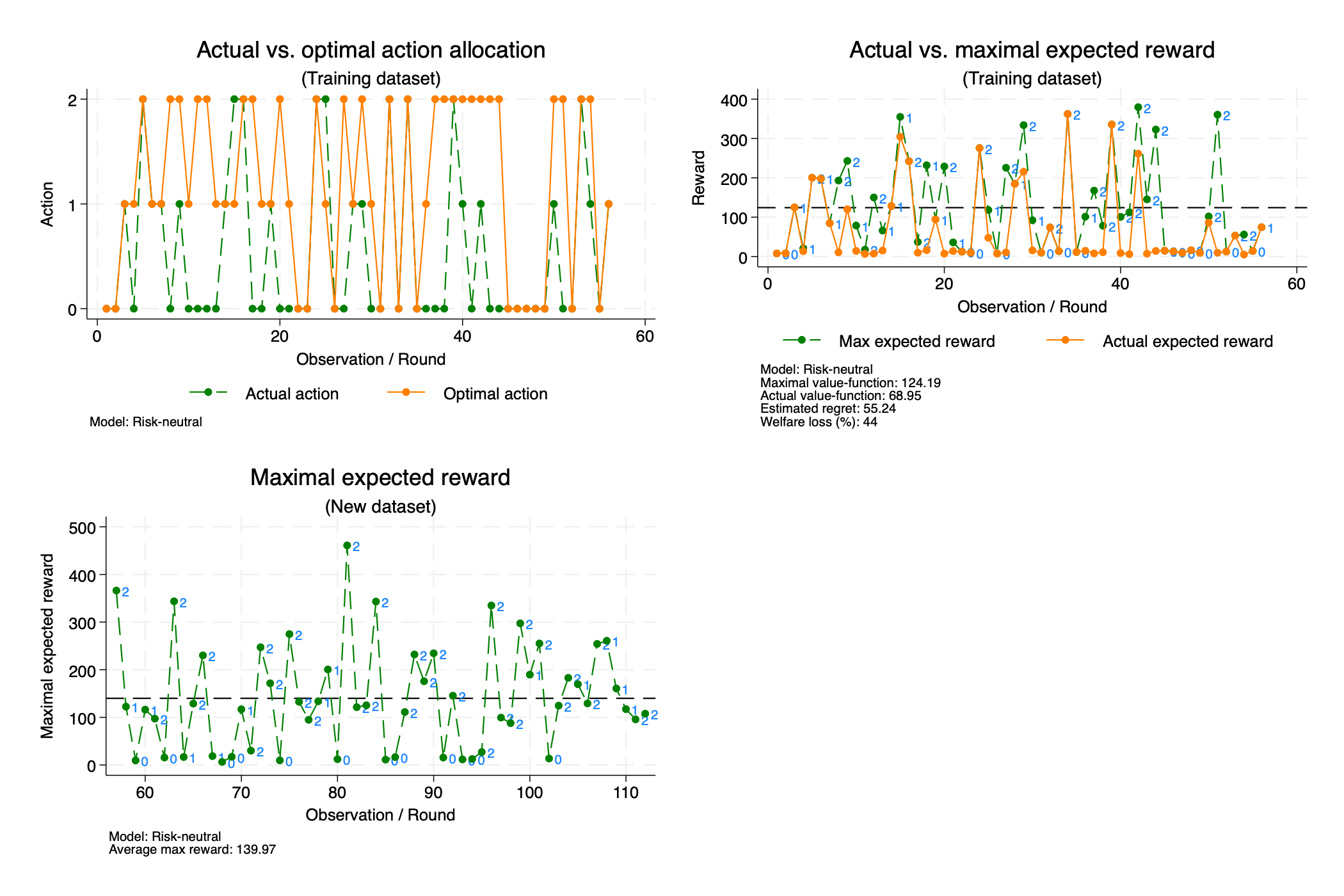}
\caption{OPL in a risk-neutral setting. According to the first-best policy rule, the figure reports: (1) actual versus optimal treatment allocation in the training data; (2) actual versus maximal expected reward in the training data; (3) maximal expected reward under the optimal treatment allocation in the new (unlabeled) observations.}
\label{fig:fig2}
\end{figure}
 
Overall, these results highlight that substantial gains in outcome value are achievable through optimal policy learning relative to random or historical policies. The learned policy can be easily projected to new data. Finally, there is a significant gap between observed treatment decisions and the model-suggested optimal policy, revealing room for improvement in actual decision-making processes.

\subsection{Value-function estimation using IPW and DR methods}

Following the estimation of the optimal policy using \texttt{opl\_ma\_fb}, we proceed to quantify the performance gap between the learned policy and the observed (historical) treatment policy using the \texttt{opl\_ma\_vf} command. This is achieved by comparing the estimated value functions of each policy under multiple estimation strategies: Regression Adjustment (RA), Inverse Probability Weighting (IPW), and Doubly Robust (DR). These approaches offer complementary perspectives on expected outcomes under different policy rules. Below, we show the code.

\begin{verbatim}
********************************************************************************
* REGRET ESTIMATION USING "opt_ma_vf"
********************************************************************************
* Value-function "first-best policy"
cap drop _D* _pi*
keep if _index==0
opl_ma_vf $y $X , policy_train($A) policy_new(_opt_policy)

-------------------------------------------------------
MAIN RESULTS
-------------------------------------------------------
--> Data information
-------------------------------------------------------
Number of training observations = 56
-------------------------------------------------------
--> Policy information
-------------------------------------------------------
Target variable: spmeasure
Features:  pindex eindex
Policy variable: w
Number of actions: 3
Actions: {0 1 2}
-------------------------------------------------------
Frequencies of the actions in the training dataset
-------------------------------------------------------

Multivalued |
 treatment: |
    j=0,1,2 |      Freq.     Percent        Cum.
------------+-----------------------------------
          0 |         34       60.71       60.71
          1 |         13       23.21       83.93
          2 |          9       16.07      100.00
------------+-----------------------------------
      Total |         56      100.00
-------------------------------------------------------
--> Value-function estimation
-------------------------------------------------------
RA = 124.19
IPW = 109.5
DR = 125.31
-------------------------------------------------------
Legend
-------------------------------------------------------
RA = Regression Adjustment
IPW = Inverse Probability Weighting
IPW = Doubly Robust
-------------------------------------------------------

gl EV_RA_opt=e(RA)    // regression adjustment
gl EV_IPW_opt=e(IPW)  // inverse probability weighting
gl EV_DR_opt=e(DR)    // double robust

* Value-function "training policy"
cap drop _D* _pi*
opl_ma_vf $y $X , policy_train($A) policy_new($A)
 
<output omitted>

-------------------------------------------------------
--> Value-function estimation
-------------------------------------------------------
RA = 68.95
IPW = 82.1
DR = 89.24
-------------------------------------------------------
Legend
-------------------------------------------------------
RA = Regression Adjustment
IPW = Inverse Probability Weighting
IPW = Doubly Robust
-------------------------------------------------------

gl EV_RA_curr=e(RA)
gl EV_IPW_curr=e(IPW)
gl EV_DR_curr=e(DR)

* Regret estimation
global regret_RA=$EV_RA_opt-$EV_RA_curr

di in red "Regret RA = "$regret_RA
* Regret RA = 55.237134

global regret_IPW=$EV_IPW_opt-$EV_IPW_curr

di in red "Regret IPW = "$regret_IPW
* Regret IPW = 27.405693

global regret_DR=$EV_DR_opt-$EV_DR_curr

di in red "Regret DR = "$regret_DR
* Regret DR = 36.070624
********************************************************************************
\end{verbatim}

The estimated value function for the learned optimal policy is consistently high across methods: 124.19 (RA), 109.50 (IPW), and 125.31 (DR). These results are fully consistent with the value reported earlier by \texttt{opl\_ma\_fb} (124.19 in training), confirming that the policy is capable of delivering substantial welfare gains, as measured by expected student performance.

In contrast, the value function associated with the actual historical assignments is markedly lower: 68.95 (RA), 82.10 (IPW), and 89.24 (DR). This reinforces the earlier insight that the treatment decisions observed in the data - despite being real-world allocations - are sub-optimal relative to what could have been achieved through data-driven decision-making.

Regret estimation, finally, is calculated as the difference between the value of the optimal policy and the value of the current (historical) policy. It quantifies the cost of suboptimal decision-making in terms of foregone outcome potential:

\begin{itemize}
\item RA-based regret: 55.24
\item IPW-based regret: 27.41
\item  DR-based regret: 36.07
\end{itemize}

These regret values are practically significant. The highest regret is seen under the RA estimator, which is consistent with the earlier results from \texttt{opl\_ma\_fb}, where the gap between optimal and observed policies was large. The DR-based regret is slightly lower but still notable, offering a robust and balanced estimate of lost welfare due to non-optimal policy choices.

The results clearly highlight that had the learned policy been implemented instead of the observed treatment allocation, average outcomes could have been substantially improved. This serves as compelling evidence for the value of OPL, especially in contexts where historical decisions were made without rigorous data-driven optimization.

Moreover, the regret computation is especially useful in policy settings, as it shifts the narrative from merely comparing average outcomes to evaluating opportunity costs of decisions. It helps policymakers understand what could have been achieved and underscores the importance of implementing better policy rules.

Together with the previous analysis, these results offer a rigorous and policy-relevant argument for adopting the OPL framework in real-world multi-action settings, such as education, healthcare, and labor policy.

The previous analysis reflects a realistic policy analysis workflow, where researchers must train models under observational assumptions and then assess the effectiveness of learned decision rules both in-sample and on new data. The careful structuring of training and evaluation phases, alongside simulated benchmarks, helps build evidence for the potential impact of policy learning in complex, multi-action settings.

\subsection{Computing policy and value function under risk aversion}

It is now interesting to learn the optimal policy (and compute the corresponding value function) when we consider a risk-averse decision maker, and then compare it with the first-best risk-neutral optimal one. 

For this purpose, we re-run \texttt{opl\_ma\_fb} by changing the \texttt{model()} option respectively to:
\begin{itemize}
\item \texttt{risk\_averse\_linear};
\item \texttt{risk\_averse\_quadratic}.
\end{itemize}
For the sake of brevity, as the changes in the code are minimal, we do not display the code, but just the main results, comparing them with the risk-neutral set-up.

However, before presenting the results, it is important to note that the conditional variance cannot be estimated perfectly, and in some cases it may even turn out negative. This represents a serious issue, as it prevents a meaningful comparison between the risk-neutral and risk-averse settings. To address this, \texttt{opl\_ma\_fb} provides the option \texttt{value\_var(\#)}, which replaces negative conditional variances with the specified value \texttt{\#}. In the example below, we set \texttt{\#} to 0.01. The results are reported below.
 
\begin{verbatim}
-------------------------------------------------------
RISK NEUTRAL
-------------------------------------------------------
--> Training data
-------------------------------------------------------
Value-function of the policy (training) = 68.95
Value-function of the non-optimal policy (training) = 57.97
Value-function of the optimal policy (training) = 124.19
Rate of optimal policy matches = .52
-------------------------------------------------------

-------------------------------------------------------
RISK AVERSE LINEAR
-------------------------------------------------------
--> Training data
-------------------------------------------------------
Value-function of the policy (training) = 61.69
Value-function of the non-optimal policy (training) = 61.04
Value-function of the optimal policy (training) = 104.16
Rate of optimal policy matches = .5
-------------------------------------------------------
Note: Negative conditional variances imputed as equal to 0.01

-------------------------------------------------------
RISK AVERSE QUADRATIC
-------------------------------------------------------
--> Training data
-------------------------------------------------------
Value-function of the policy (training) = 61.49
Value-function of the non-optimal policy (training) = 63.06
Value-function of the optimal policy (training) = 88.15
Rate of optimal policy matches = .48

-------------------------------------------------------
Note: Negative conditional variances imputed as equal to 0.01
\end{verbatim}

Results are significantly different, although the rate of optimal policy matches is similar, around 50\%. The value function of the first-best risk neutral decision rule is equal to 124.19. As expected, the value function of the linear risk-aversion decision rule is smaller and equal to 104.16, which produces a regret of $20.03 = 124.19 - 104.16$. For the quadratic risk-aversion decision rule, the value function is even smaller and equal to 88.15, producing a regret of $36.04 = 124.19 - 88.15$.

In some cases, variance imputation may be unreliable.
As an alternative, one can compare the risk-neutral and risk-averse settings by restricting the analysis to cases where the conditional variance is nonnegative in both settings. For this purpose, I developed two simple utility commands, \texttt{opl\_best\_treat} and \texttt{opl\_plot\_best}. These commands are used after estimating \texttt{opl\_ma\_fb} and loading the dataset that stores the estimated conditional predictions and variances, specified in the \texttt{save\_preds\_vars(name)} option, where \texttt{name} is the name of the dataset. Below a Stata code doing this.

\begin{verbatim}
***************************************************************************
* Restricting value function estimation to cases with 
* nonnegative conditional variances
***************************************************************************

* Load the data
use mydata , clear                        

* Define global macro "model" = "risk_neutral"
global model "risk_neutral"               

* Run OPL estimator (multi-action, first-best)
qui opl_ma_fb $y $X, policy_train($A) ///
    model($model) ///
    name_opt_policy("_opt_policy") ///
    match_name("_match_var") ///
    new_data(my_new_data) ///
    gr_action_train("gr1") ///
    gr_reward_train("gr2") ///
    gr_reward_new("gr3") ///
    save_preds_vars("my_results") ///
    policy_non_optimal_train(treat_non_optimal_train)

<output omitted>

* Compute the first-best solution for "risk neutral" by deleting rows 
* with negative conditional variances (so comparison with risk-averse 
* setting is coherent, without variance imputation).

* Load the saved prediction results
use my_results , clear                    

* Generate a variable with 1 = missing values across __var* variables
egen nmiss = rowmiss(__var*)              

* Drop rows with missing values (negative variance cases)
drop if nmiss > 0                         

* Keep only training observations (_index=0).  
* Alternatively, could keep new obs (_index=1), or both.
keep if _index==0                         

* Assume w ranges from 0 to M=3
local M = 3                               

* Initialize variable for selected prediction
gen __pred_sel = .                        

* Loop over treatments j=0,...,M-1
* For each obs, pick the prediction corresponding to its actual treatment w
forvalues j = 0/`=`M'-1' {                
    replace __pred_sel = __pred`j' if w == `j'
}

* Compute best treatment across predicted outcomes using "opl_best_treat"
opl_best_treat __pred0 __pred1 __pred2        

* Summarize maximum predicted outcomes
sum __Y_hat_max                           

* Display the average value function
di "Value function risk neutral (with missing for negative vars) = " r(mean) 
196.23337

* Plot observed vs predicted best treatments using "opl_plot_best"
* Order: Y_hat_obs, T_obs, Y_hat_max, T_best
* Also produce graphs of predicted rewards
opl_plot_best __pred_sel w __Y_hat_max __T_best , ///
gr_reward_train(Graph_reward) gr_action_train(Graph_action)        
***************************************************************************                            
\end{verbatim}

Under this restriction, the value function in the risk-neutral setting is 196.23. In comparison, the value function in the risk-averse linear setting without variance imputation is 190.78 (code omitted for brevity). This results in a substantially smaller regret, equal to $5.45 = 196.23 - 190.78$.

\begin{figure}[t]
\centering
\includegraphics[width=12cm]{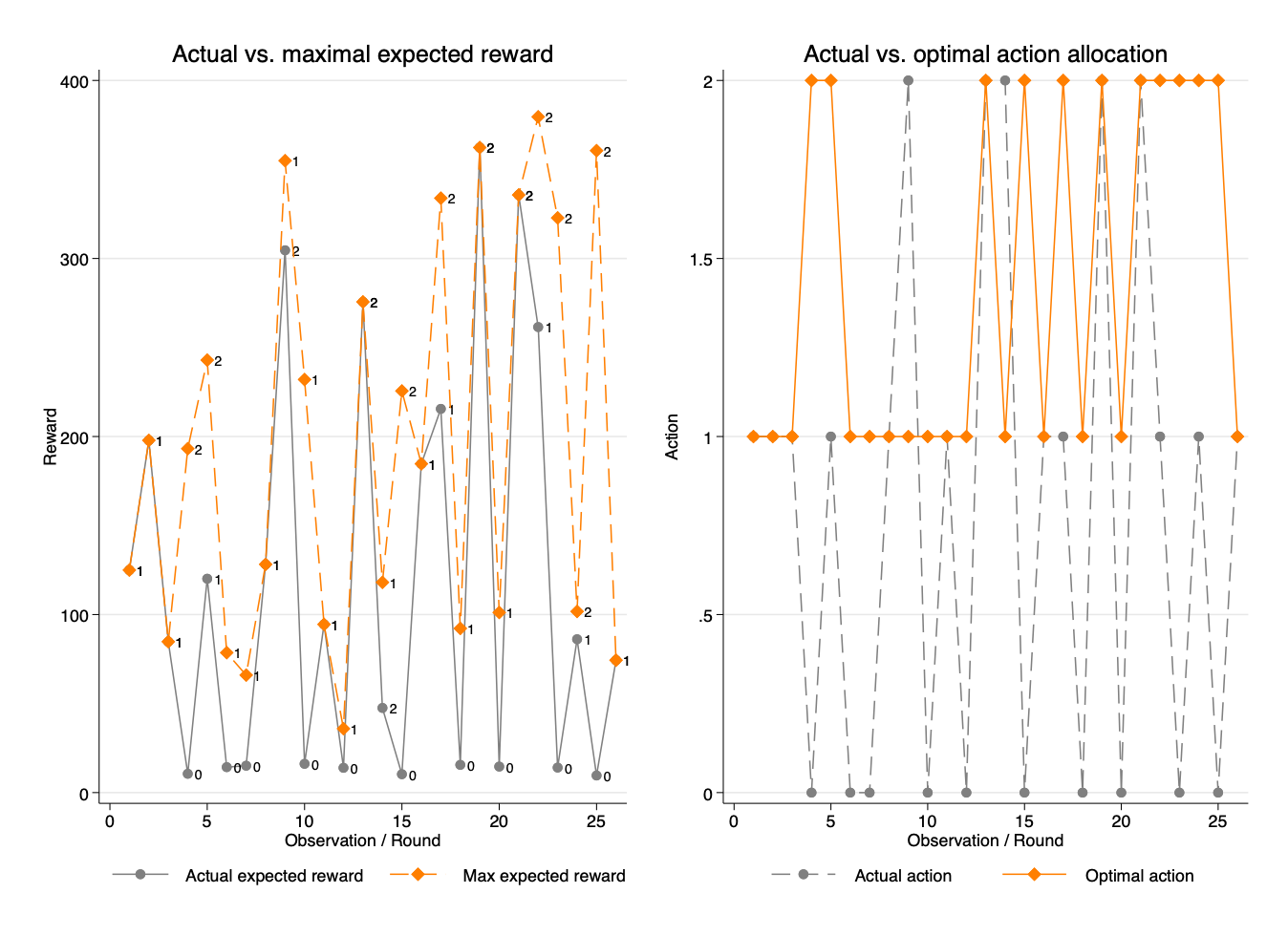}
\caption{OPL in a risk-neutral setting, by eliminating observations with negative conditional variances. Graphs obtaining by running the \texttt{opl\_plot\_best} command after \texttt{opl\_ma\_fb} and then \texttt{opl\_best\_treat}. Left: actual versus maximal expected reward in the training data. Right: actual versus optimal treatment allocation in the training data.}
\label{fig:GraphFinal}
\end{figure}

Figure \ref{fig:GraphFinal} shows the main output of the command \texttt{opl\_plot\_best} run after \texttt{opl\_ma\_fb} and then \texttt{opl\_best\_treat}. The figure reports: (1) actual versus maximal expected reward in the training data; (2) actual versus optimal treatment allocation in the training data. This allows for comparing risk-neutral and risk-averse settings in the presence of negative conditional variances.

\subsection{Regret interpretation}

The values of the regrets under risk-averse preferences can be interpreted as a measure of the \textit{social cost} of incorporating risk considerations into treatment allocation decisions. Such costs can only be evaluated ex post, once decisions have already been made, but they may also help explain why the first-best risk-neutral decision rule is rarely implemented in real-world settings.

Put differently, accounting for risk can be viewed as a \textit{constraint} that policymakers may incorporate when learning optimal policies. Risk, however, represents only one among many possible constraints that arise in practice, alongside ethical, legal, or equity considerations. The presence of such policy constraints typically leads to the adoption of rules that are suboptimal in terms of average reward, but that protect the decision maker against unfavorable outcomes. To illustrate this point more clearly, I present a simple simulation example.   

\noindent
Consider an individual $i$, characterized by features $X_i$. Given $X_i$, the policymaker must decide whether to treat $i$ or not.  
Suppose that, conditional on $X_i$, individual $i$ has the following potential outcome distributions (for simplicity, in what follows, I suppress the index $i$):

\begin{equation}
Y_1|X \sim \mathcal{N}(75, 65), \qquad 
Y_0|X \sim \mathcal{N}(30, 10).
\label{eq:sim1}
\end{equation}
This setup simulates a situation in which treatment yields a higher expected reward than non-treatment, but at the cost of much greater variance.
  
According to the first-best rule, individual $i$ should always be treated, regardless of outcome variance. Indeed:
\[
\tau(X) = m_1(X) - m_0(X) = 75 - 30 = 45 > 0,
\]
where $m_1(X) = E(Y_1|X) = 75$ and $m_0(X) = E(Y_0|X) = 30$, which implies that treatment is optimal.  

This corresponds to the risk-neutral case, in which the policymaker considers only expected outcomes and chooses the treatment with the higher mean reward. However, the first-best solution may differ from the \textit{oracle} solution, which accounts for the entire distributions of $Y_1|X$ and $Y_0|X$.  
The oracle rule treats $i$ if $Y_1|X \geq Y_0|X$, and does not treat otherwise. Yet, due to randomness, $Y_1|X \neq m_1(X)$ and $Y_0|X \neq m_0(X)$.  

In general, high variability may induce \textit{treatment inversions}: the first-best rule may recommend treatment when it is actually harmful, or non-treatment when treatment would be beneficial. The frequency of such inversions depends on the relative variances of the two distributions. In this example, we explore the case where the variance of $Y_1|X$ is much higher than that of $Y_0|X$.  

We simulate 10{,}000 draws from the distributions in (\ref{eq:sim1}) and compute:
\[
\text{(i) Risk-neutral first-best policy: } \mathbbm{1}[m_1(X) - m_0(X) > 0],
\]
\[
\text{(ii) Oracle policy: } \mathbbm{1}[Y_1|X - Y_0|X > 0],
\]
\[
\text{(iii) Risk-averse linear first-best policy: } \mathbbm{1}\!\left[\frac{m_1(X)}{s_1(X)} > \frac{m_0(X)}{s_0(X)}\right].
\]
\noindent
We then compare, for the same individual, realized outcomes and expected value functions under the risk-neutral and risk-averse settings when risk is high.\footnote{The Stata code used in this analysis is available from the author upon request and will be made publicly accessible.} 

\begin{figure}[t]
\centering
\includegraphics[width=12cm]{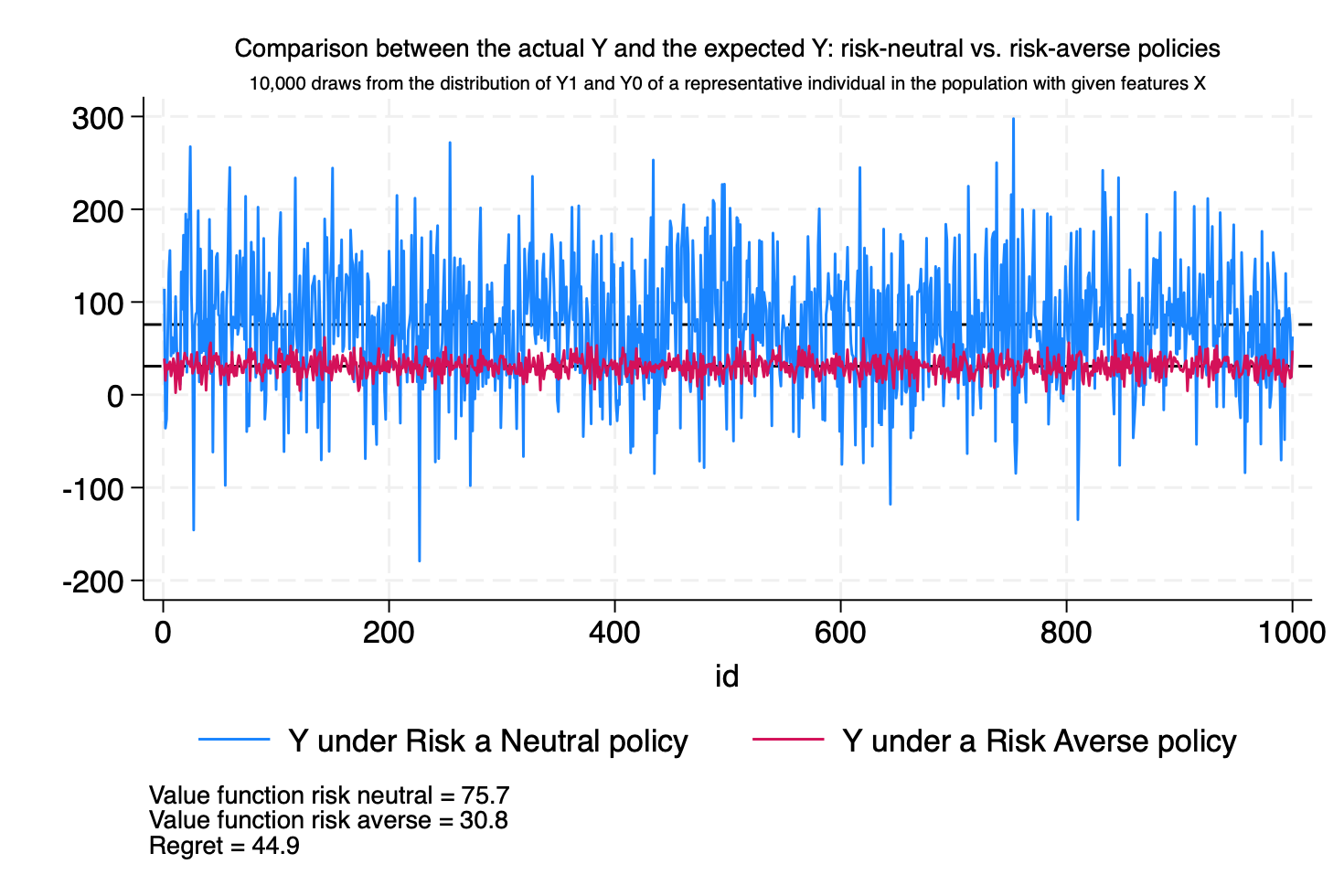}
\caption{Realized outcomes under risk-neutral (blue) and risk-averse (red) policies. The risk-neutral policy yields higher expected welfare but with much greater variability; the risk-averse policy reduces dispersion at the cost of lower welfare.}
\label{fig:fig3}
\end{figure}

A key insight emerges: while the value functions (average rewards) are similar across settings, the realized outcomes differ substantially. Under the risk-neutral policy, individual $i$ faces a high probability of extreme values (very large or very small $Y$). Under the risk-averse policy, the outcome distribution is more concentrated, reducing exposure to extremes. Figure \ref{fig:fig3} illustrates this result.    

In the worst-case scenario, where all individuals realize their minimum outcomes, the risk-neutral policy produces very low average welfare, whereas the risk-averse policy delivers much higher average welfare. Conversely, in the best-case scenario, where all individuals realize their maximum outcomes, the risk-neutral policy vastly outperforms the risk-averse one, since it fully exploits the upper tail of the distribution of $Y_1|X$.  

A policymaker does not know in advance which scenario will occur. To avoid potentially severe losses in the worst case, she may prefer to sacrifice some average welfare. Risk aversion, therefore, implies an \textit{average social cost} (or \textit{average regret}), which is the price paid to insure against unfavorable outcomes.  

Ultimately, the choice of policy is not just about adopting the universally best strategy, but about the decision maker's attitude toward risk. This attitude is highly context-dependent and may vary substantially across policy domains.

\section{Conclusion}
\label{sec:conclusion} 

This paper introduces a comprehensive Stata implementation for optimal policy learning in multi-action treatment settings, explicitly accounting for different risk preferences on the part of the policymaker. The two companion commands, \texttt{opl\_ma\_fb} and \texttt{opl\_ma\_vf}, provide researchers with practical tools to estimate optimal policy assignments from observational data, bridging advanced econometric theory with real-world applications.

The core innovation of this work lies in making complex value-based policy learning accessible through the familiar environment of Stata. Users can specify risk-neutral, linear risk-averse, or quadratic risk-averse preferences, allowing for flexible modeling of decision-makers’ attitudes toward outcome variability. The commands are designed to be intuitive and computationally efficient, facilitating widespread use in policy evaluation across disciplines.

The paper has also provided a clear formalization of the value function and its role in defining optimal policies under uncertainty. Special attention has been given to the challenges of estimating counterfactual outcomes from observational data and to the practical implications of adopting different risk specifications.

The empirical application has demonstrated how the proposed tools can be used to derive policy rules tailored to individual observed characteristics, highlighting their effectiveness in guiding individualized, data-driven decision-making.

Future research directions include extending the methodology beyond the first-best optimal rule, typically unconstrained and unstructured, by incorporating realistic policy classes such as threshold-based decision rules, linear index policies, and regression tree-based strategies. These structured policy classes are often more interpretable, implementable, and better aligned with real-world policy design. 

The proposed commands estimate conditional means using a linear regression model. A potential improvement would be to incorporate supervised machine learning techniques for this task, such as tree-based models (including random forests and boosting), regularized approaches (such as lasso and elasticnet), or even neural networks. 

Finally, we plan to develop extensions that accommodate fairness, equity, feasibility, and budgetary constraints into the optimization process.

\section{Acknowledgment}
\label{sec:Acknowledgment} 
This work was supported by: FOSSR (Fostering Open Science in Social Science Research),
funded by the European Union - NextGenerationEU under the NPRR grant agreement MUR IR0000008;
PRIN Project RECIPE (Linking Research Evidence to Policy Impact and Learning: Increasing the Effectiveness of Rural Development Programmes Towards Green Deal Goals), MUR code: 20224ZHNXE. 

\newpage
%

\bibliographystyle{sj}
\nocite*{}
\bibliographystyle{chicago}


\begin{aboutauthors}
Giovanni Cerulli is a senior researcher at the CNR-IRCrES, Research Institute on Sustainable
Economic Growth, National Research Council of Italy, Rome. His research interest is in applied
econometrics, with a special focus on causal inference and machine learning. He has developed
original causal inference models and provided several implementations. He is currently editor
in chief of the \textit{International Journal of Computational Economics and Econometrics}.
\end{aboutauthors}

\clearpage
\end{document}